\documentclass[runningheads]{llncs}

\usepackage{url}
\usepackage{hyperref}
\usepackage{siunitx}
\usepackage{multirow} 
\usepackage{setspace}
\usepackage{blindtext}
\usepackage{scrextend}
\usepackage{graphicx}
\usepackage{caption}
\usepackage[font=footnotesize,labelformat=simple]{subcaption}
\usepackage{subfloat}
\usepackage{float}
\usepackage{enumerate}
\usepackage{amsmath}
\usepackage{color}
\usepackage{CJKutf8}
\usepackage{algorithmic}
\usepackage{algorithm}
\usepackage{tabularx}
\usepackage{booktabs}
\usepackage{tcolorbox}
\usepackage{lmodern}

\graphicspath{{images/}}
\begin{document}
\title{Leveraging Hardware Performance Counters for Predicting Workload Interference in Vector Supercomputers}
\titlerunning{Leveraging Hardware Performance Counters}
%
\author{
Shubham\inst{1} \and
Keichi Takahashi\inst{2} \and
Hiroyuki Takizawa\inst{2} 
}

%
\authorrunning{Shubham \textit{et al.}}
%
\institute{
Graduate School of Information Sciences, Tohoku University\\
\email{shubham.s4@dc.tohoku.ac.jp}\\ \and
Cyberscience Center, Tohoku University\\
\email{\{keichi,takizawa\}@tohoku.ac.jp} 
}
\maketitle              
\begin{abstract}
In the rapidly evolving domain of high-performance computing (HPC), heterogeneous architectures such as the SX-Aurora TSUBASA (SX-AT) system architecture, which integrate diverse processor types, present both opportunities and challenges for optimizing resource utilization. This paper investigates workload interference within an SX-AT system, with a specific focus on resource contention between Vector Hosts (VHs) and Vector Engines (VEs). Through comprehensive empirical analysis, the study identifies key factors contributing to performance degradation, such as cache and memory bandwidth contention, when jobs with varying computational demands share resources. To address these issues, we develop a predictive model that leverages hardware performance counters (HCs) and machine learning (ML) algorithms to classify and predict workload interference. Our results demonstrate that the model accurately forecasts performance degradation, offering valuable insights for future research on optimizing job scheduling and resource allocation. This approach highlights the importance of adaptive resource management strategies in maintaining system efficiency and provides a foundation for future enhancements in heterogeneous supercomputing environments.

\end{abstract}
\section{Introduction}
HPC systems have become the backbone of scientific research and technological innovation, enabling the processing of complex computations across various fields such as climate modeling, molecular dynamics, and data analytics. In recent years, there has been a significant shift towards heterogeneous architectures within HPC systems, combining different types of processors to maximize computational efficiency. These architectures typically integrate Central Processing Units (CPUs) with specialized accelerators such as Graphics Processing Units (GPUs) or VEs, creating a diverse ecosystem of computational resources.

\begin{figure}[tbp]
{
  \begin{center}
    \includegraphics[width=\textwidth]{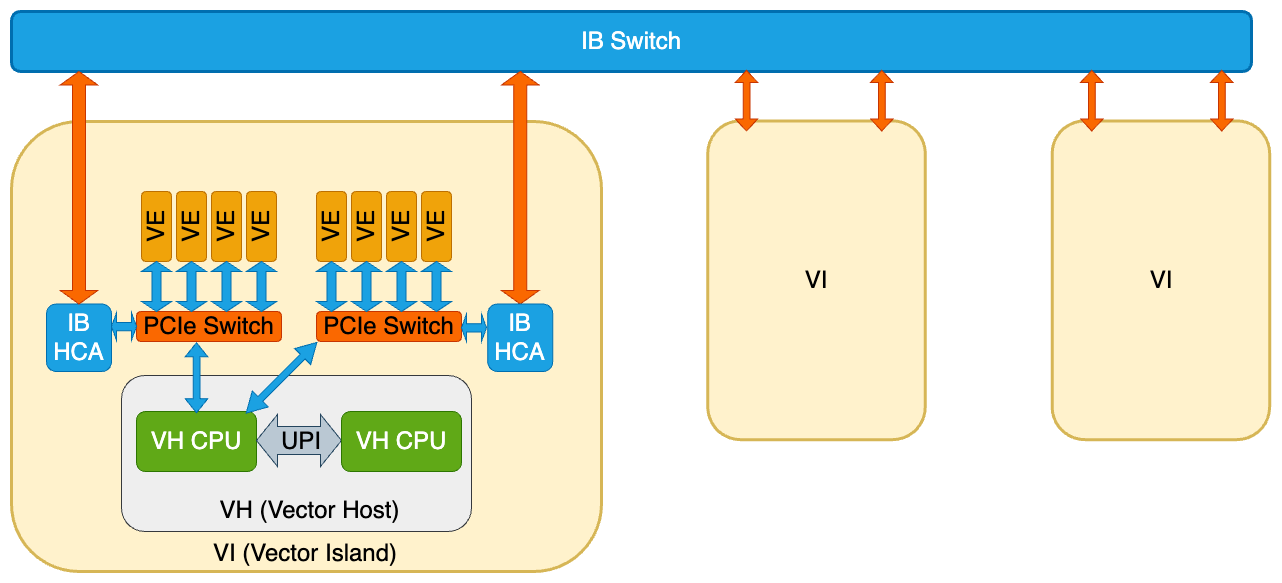}
    \caption{Architecture of multiple VI system~\cite{sx}.}
    \label{fig:vi}
  \end{center}
}
\vspace{-0.8cm}
\end{figure}

One of the leading examples of such heterogeneous systems is the SX-AT supercomputer~\cite{Komatsu,yamada}, developed by NEC. The SX-AT system embodies a sophisticated heterogeneous architecture, featuring both VHs and VEs. The VHs, based on standard x86 processors, manage the operating system and oversee the VEs, which are designed for memory-intensive scientific calculations and are implemented as PCI-Express cards with high memory bandwidth. A single compute node in the SX-AT system, known as a Vector Island (VI), combines multiple VHs and VEs as seen in Figure~\ref{fig:vi}.

The impact of resource contention and workload interference on HPC systems like the SX-AT is significant. In the SX-AT system, VH workloads handle general-purpose computations, while VE workloads focus on memory-intensive scientific calculations. However, VE workloads still depend on the VH for system operations, such as system calls, which are forwarded to the VH within the same VI for processing by the operating system. This reliance on the VH means that frequent system calls from the VE consume VH CPU time and shared resources like network bandwidth, leading to resource contention. Performance degradation occurs when computationally intensive VH workloads run simultaneously with VE workloads that frequently generate system calls, as the VH must juggle both its own workload and the system calls of VE. This interference underscores the challenge of efficiently managing heterogeneous resources and highlights the need for advanced resource management strategies to mitigate performance bottlenecks.

To address these challenges, this paper presents a predictive model that leverages HCs and applies ML techniques to predict the likelihood of performance degradation due to VH-VE workload interference. HCs provide valuable insights into system behavior by monitoring resource usage such as CPU cycles, cache utilization, and memory bandwidth. These metrics serve as input features for the ML model, which is designed to classify and predict workload pairs that are likely to cause significant resource contention. By identifying interference-prone workload combinations, the model can guide efficient job scheduling and resource allocation, ultimately improving the overall utilization and performance of the SX-AT system.
\begin{figure}[tp]
\centering
\includegraphics [width=\textwidth] {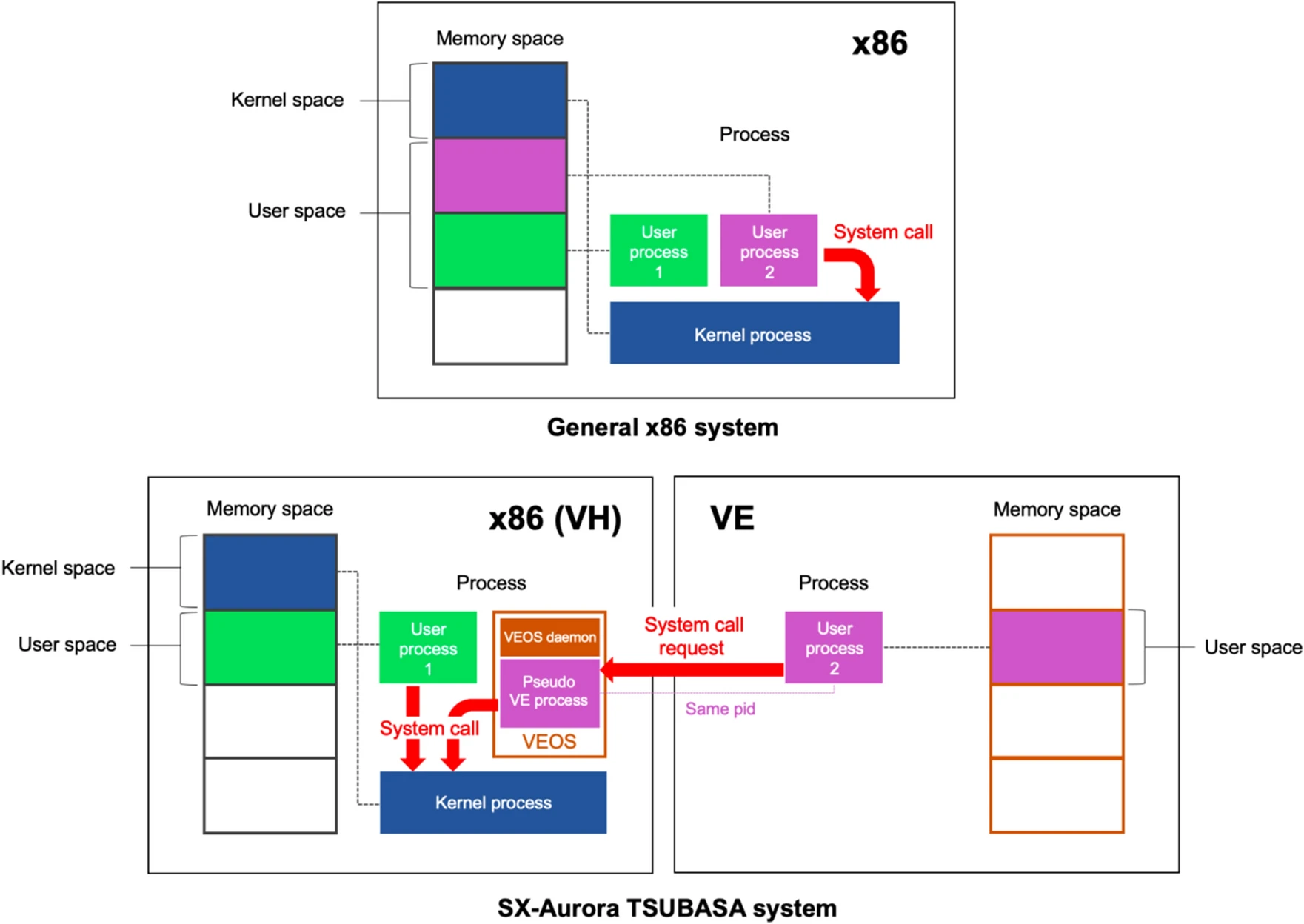}
\caption{System Architecture and Process Interaction in SX-Aurora TSUBASA Systems~\cite{nuno}.}
\label{fig:arc}
\vspace{-0.57cm} 
\end{figure}
The main objective of this research is to enhance the efficiency of the SX-AT system by developing a reliable prediction model for workload interference. This study not only contributes to the theoretical understanding of workload interference in SX-AT systems but also offers practical solutions for improving the efficiency and effectiveness of supercomputing infrastructures.

\section{Conflict Mechanism in SX-Aurora TSUBASA System}

In the SX-AT system, conflicts arise due to the shared usage of computing resources between VHs and VEs. The system consists of VHs and VE components, where the user processes on the VE access memory spaces located on VE memory devices, which benefit from high memory bandwidth. However, kernel memory space resides on the VH, even when an application runs on a VE. As illustrated in Figure~\ref{fig:arc}, when a VE process requires VH resources, it sends a system call request to a pseudo-VE process on the VH, and the VEOS daemon on the VH forwards it to the kernel, creating a dependency on VH resources. Both VHs and VEs share resources such as CPU time, memory bandwidth, network, and file access. When the VE workload initiates system calls, the VH workload would be context-switched to the pseudo-VE process so that the VH core can handle the system call request, potentially causing a delay in VH workload execution. Additionally, intensive usage of shared resources by either VH or VE workload can lead to performance degradation due to resource contention. For instance, memory-intensive VE 
workloads on a VH can consume memory bandwidth, increasing memory access latency for VE processes. Similarly, file system access conflicts can further increase execution time if both workloads frequently access the file system. To mitigate this performance degradation, it is crucial to identify application characteristics that cause conflicts and to optimize job scheduling, avoiding resource contention and improving overall system performance.

\section{Related Work}
The performance evaluation and application of SX-AT in scientific computations have been extensively documented by various researchers. Takahashi et al.~\cite{takahashi} have explored the potential of SX-AT in different scientific domains, highlighting its computational efficiency and effectiveness. Additionally, the VH-VE offload programming model has been examined in depth, with papers such as~\cite{ke} and~\cite{takizawa} focusing on the programming methodologies and their impacts on performance. I/O performance enhancements and challenges associated with SX-AT have also been investigated~\cite{yuta}.

Despite these comprehensive studies, there is a notable gap in the literature regarding the quantitative evaluation of performance interference when VHs and VEs workloads coexist. This gap was partially addressed by~\cite{nuno}, which analyzed system throughput by co-executing workloads on both VH and VE. Their study demonstrated that the system call frequency from VE workloads could predict performance degradation, suggesting that using the CPU load of VH as an approximation for system call frequency could aid in identifying conflicting workload pairs at runtime.
\begin{figure}[tp]
\centering
\includegraphics[width=\textwidth]{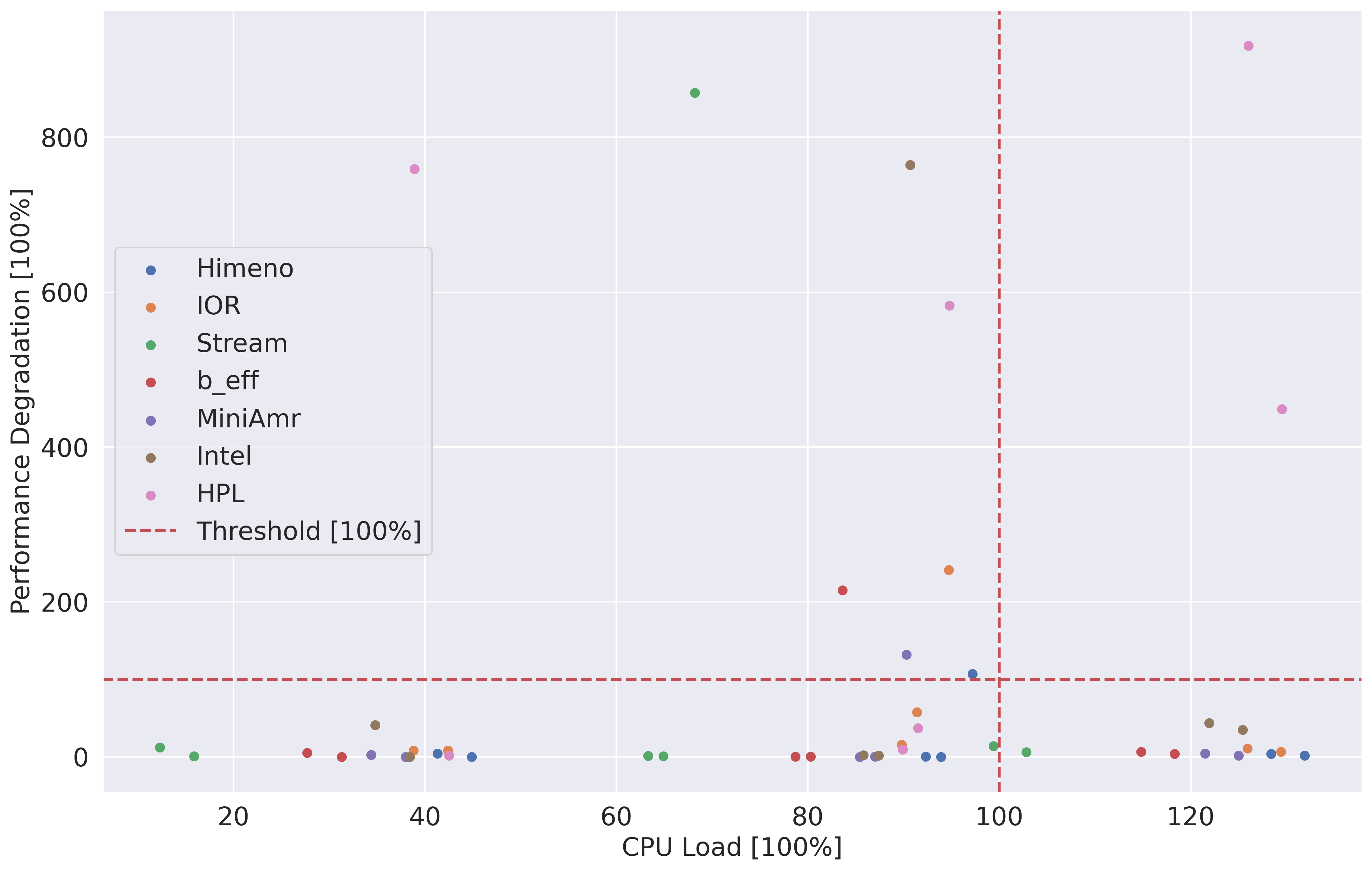}
\caption{Interference Comparison between CPU load vs Performance Degradation.}
\label{fig:interference} 
\vspace{-1.18em} 
\end{figure}
However, the approach taken by~\cite{nuno} has several limitations. Firstly, the assumption of a fixed execution time of 100 seconds for every workload does not accurately reflect real-world scenarios, where workloads typically have varying execution times. Secondly, their conflict model primarily considers CPU load as the sole metric for predicting interference. This simplification potentially overlooks other critical factors, such as memory bandwidth contention and cache conflicts, which can also significantly impact performance.

As illustrated in Figure~\ref{fig:interference} (scatter plot of CPU load vs performance degradation), some workloads experience high interference even when their CPU loads are below 100\%. The threshold of 100\% used in~\cite{nuno} is not always reliable, as some workloads exhibit low performance degradation despite having a CPU load exceeding 100\%. This discrepancy indicates that the CPU load technique does not consistently classify workloads into high and low interference categories, highlighting the need for more nuanced and advanced features for classification. To address these limitations, the proposed approach in this study involves the collection of HC values to capture a broader range of factors, including cache usage and memory bandwidth. By developing an ML algorithm that incorporates these diverse metrics, this approach aims to accurately predict performance interference for various types of workloads, rather than relying on a fixed execution time. This approach is designed to be more adaptable to real-world conditions and more effective in larger, more complex systems with diverse workloads and higher degrees of parallelism.

\section{Performance Interference in SX-AT by Workload Co-execution}
\subsection{Performance Interference Metric }
To quantify performance interference, the following metric is used:
\begin{equation}
    Pd_{ve,vh} = \left( \frac{T_{co\text{-}location_{ve,vh}} - T_{solo_{ve}}}{T_{solo_{ve}}} \right) \times 100,
\end{equation}
where ${T_{co\text{-}location_{ve,vh}}}$ represents the execution time of the VE workload when it is running concurrently with the VH workload, and $T_{solo_{ve}}$ represents the execution time of the VE workload when it is running in isolation. $ Pd_{ve,vh}$ denotes the performance degradation of VE workload running in a collocated manner in VEs and VHs, and this metric provides a percentage value that indicates the relative increase in execution time due to the presence of the interfering workloads. A higher value indicates a greater performance degradation.

\subsection{Benchmarks}
\begin{table}[h]
    \centering
    \caption{List of Benchmarks.}
    \begin{tabular}{ll}
        \toprule
        \textbf{Benchmark} & \textbf{Version} \\
        \midrule
        Himeno benchmark~\cite{himeno} & 3.0 \\
        Interleaved or Random (IOR) benchmark~\cite{ior} & 4.1 \\
        Intel MPI benchmark~\cite{intel} & 2021.7 \\
        STREAM Benchmark~\cite{stream} & 5.10 \\
        Effective Bandwidth (b\_eff) Benchmark~\cite{b_eff} & 3.6.0.1 \\
        MiniAMR benchmark~\cite{mini} & N/A \\
        High-Performance Linpack (HPL) benchmark~\cite{hpl} & 2.3 \\
        \bottomrule
    \end{tabular}
    
    \label{table:benchmarks}
    \vspace{-1.18em} 
\end{table}

This study utilizes a diverse set of HPC benchmarks to assess system performance and workload interference in the SX-AT system. The Himeno benchmark evaluates computational efficiency for solving the Poisson equation, focusing on memory bandwidth. The IOR benchmark tests parallel file system performance under various I/O patterns, while the Intel MPI benchmark measures Message Passing Interface operations. The STREAM benchmark evaluates sustainable memory bandwidth, and b\_eff assesses communication bandwidth in parallel systems. The MiniAMR benchmark focuses on Adaptive Mesh Refinement algorithms, and HPL (High-Performance Linpack) evaluates floating-point computational power.

As listed in Table~\ref{table:benchmarks}, these benchmarks represent a broad spectrum of system resource utilization, including memory, I/O, and computation, offering a comprehensive evaluation of workload behavior under different conditions. This enables a detailed analysis of how these resources are affected by workload interference in the SX-AT system.

The specifications of the system used in this study are summarized in Table~\ref{table:hardware_config}. The benchmarks are compiled for both VH and VE and run using all cores on each processor.
\vspace{-0.7cm}
\begin{table}[h]
\centering
\caption{Hardware and Software configuration of NEC SX-Aurora TSUBASA.}
\begin{tabular}{ll}
\hline
\textbf{Component} & \textbf{Configuration}                   \\ \hline
Vector Host        & Intel Xeon Silver 4208 (16 cores)              \\
Vector Engine      & NEC Vector Engine Type 20B (8 cores)                       \\
Operating System   & Rocky Linux release 8.8 (Green Obsidian) \\
VEOS               & veos-3.2.1-1.el8.x86\_64                 \\
VH Compiler        & gcc-8.5.0-18.el8.x86\_64                 \\
VE Compiler        & ncc-5.1.0                                \\ \hline
\end{tabular}%

\label{table:hardware_config}
\vspace{-0.7cm}
\end{table}

\subsection{Interference Analysis}
\begin{figure}[h]
\vspace{0.2cm}
	\centering
	\includegraphics [width=\textwidth] {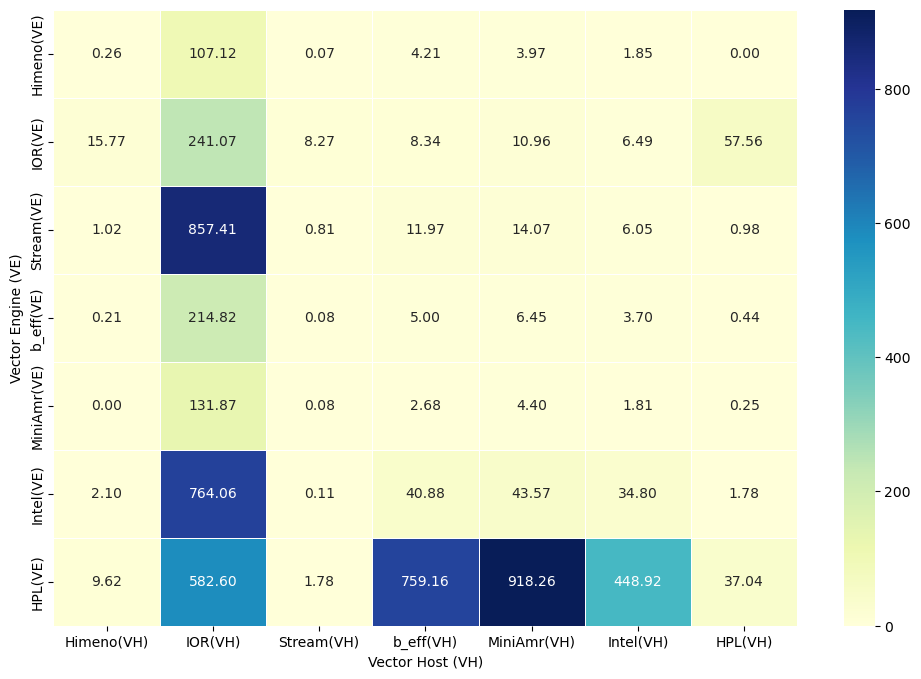}
	\caption{Performance Degradation in \% of Benchmark Running on VE and VH. }
	\label{fig:pd}
 \vspace{-0.6cm}
\end{figure}

Figure~\ref{fig:pd} illustrates the performance degradation of all seven benchmarks when executed concurrently on the VEs and VHs, resulting in 49 combinations. The primary application is run on VEs, while the secondary application is run on VHs.

Himeno on VEs generally shows minimal disruption when paired with various VH benchmarks. However, significant degradation occurs when Himeno on VEs runs concurrently with IOR on VHs, with a value of 107.12\%. This is attributed to the dependency of Himeno on memory bandwidth, which causes contention when IOR requires substantial I/O operations, leading to increased memory access latency.

STREAM on VEs exhibits an extremely high degradation of 857.41\% when paired with IOR on VHs, indicating substantial competition for memory bandwidth. Both STREAM and IOR demand substantial memory bandwidth, causing severe contention and performance degradation. Similarly, b\_eff on VEs shows moderate degradation of 214.82\% with IOR on VH, due to communication-intensive operations b\_eff interfering with I/O operations of IOR, leading to increased latency.
Intel MPI on VEs experiences a high degradation of 764.06\% when run with IOR on VHs. MPI operations are disrupted by I/O operations of IOR, resulting in significant performance bottlenecks. HPL on VEs also faces significant degradation of 448.92\% when paired with Intel MPI on VHs, due to the computational intensity of HPL creating bottlenecks.

When HPL on VEs is paired with MiniAMR on VHs, severe degradation of 918.26\% occurs. The high demand of HPL for computational resources and memory bandwidth causes substantial contention, exacerbated by similar resource requirements of MiniAMR.

The primary factors responsible for performance degradation include memory bandwidth contention, I/O operations, and computational resource contention. Benchmarks such as Himeno, STREAM, MiniAMR, and HPL heavily utilize memory bandwidth, leading to high degradation when paired with other memory-intensive or I/O-intensive benchmarks such as IOR. Intensive file system operations of MiniAMR significantly degrade performance when running with benchmarks requiring substantial memory or computational resources. Computationally intensive benchmarks such as HPL cause performance degradation due to their high demand for CPU time and memory, leading to contention with other benchmarks requiring similar resources.

In conclusion, the performance degradation observed in these experiments highlights the importance of understanding the resource demands of concurrent benchmarks. Memory bandwidth contention, intensive I/O operations, and high computational resource demands are critical factors leading to performance degradation.
\section{Workload Interference Prediction Model}
\vspace{-0.1cm}
\subsection{Data Collection}
We collected data from the SX-AT system using a suite of benchmarks, including the Himeno, IOR, and STREAM benchmarks. Each benchmark was executed in isolation and in a collocated manner on VEs and VHs to measure the extent of performance degradation due to interference. We employed an Intel Performance Counter Monitor (IntelPCM)~\cite{pcm} to collect hardware counter metrics during these executions. Specifically, while running VH and VE workloads in parallel, we repeatedly executed the workloads that were completed first until the longer-running workload was finished, ensuring comprehensive data collection. The collected metrics include Instructions Per Cycle (IPC), cache miss rates (L2 and L3 cache misses per instruction), memory bandwidth usage (read and write traffic), and other critical indicators of resource usage and performance as summarized in Table~\ref{tab:metrics}. These metrics provide a detailed profile of system behavior under varying interference conditions, enabling us to quantify the extent of performance degradation caused by resource contention.

\begin{table}
\centering
\caption{Metrics available for profiling in PCM.}
\begin{tabular}{@{}lll@{}}
\toprule
\textbf{Field} & \textbf{Explanation}                     & \textbf{} \\ \midrule
EXEC           & Instructions per nominal CPU cycle       &           \\
IPC            & Instructions per cycle (core efficiency) &           \\
L3MISS         & L3 Cache line misses (millions)          &           \\
L2MISS         & L2 Cache line misses (millions)          &           \\
L3HIT          & L3 Cache hit ratio (hits/reference)      &           \\
L2HIT          & L2 Cache hit ratio (hits/reference)      &           \\
L3MPI          & L3 Cache misses per instruction          &           \\
L2MPI          & L2MISS per instruction          &           \\
READ           & DRAM Memory read traffic in GB           &           \\
WRITE          & DRAM Memory write traffic in GB          &           \\ \bottomrule
\end{tabular}%

\label{tab:metrics}
\end{table}
\begin{figure}[h]
	\centering
	\includegraphics [width=\textwidth] {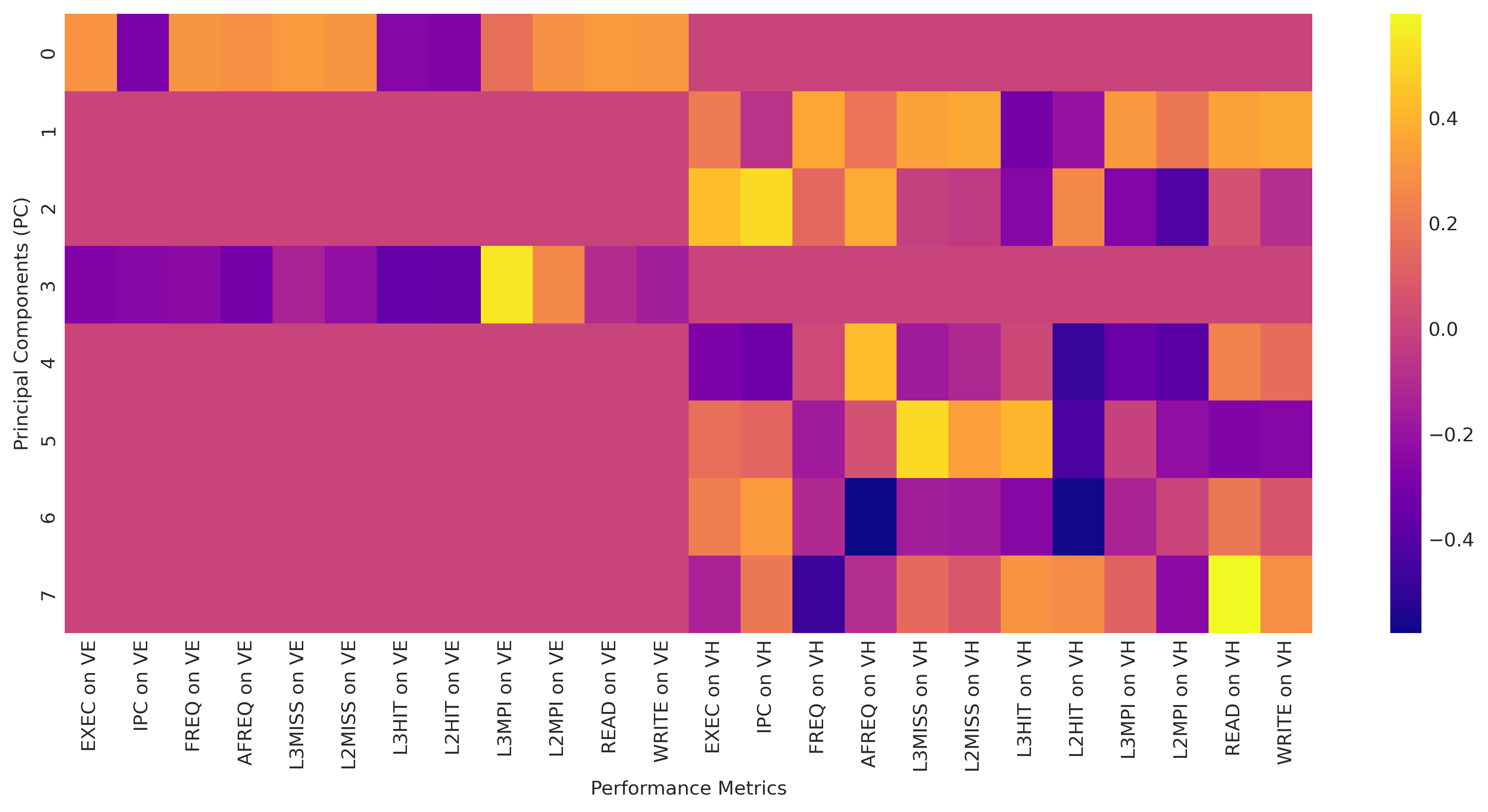}
        \vspace{-0.3cm} 
	\caption{ Principal Component Loadings for Performance Metrics on VEs and VHs. }
	\label{fig:pca}
 \vspace{-1.0em} 
 
\end{figure}

\subsection{Feature Engineering}

After collecting a dataset with 4,900 samples and 1,000 features, feature engineering was performed to enhance model training. Initially, we created a correction matrix to reveal high multicollinearity among features while eliminating data with NaN values. Principal Component Analysis was applied to reduce dimensionality, retaining eight principal components that explained 98.64\% of the total variance. Features with high absolute loadings, such as EXEC, FREQ, IPC, and L3MISS on VEs and VHs, were selected for their significant contributions as shown in Table~\ref{tab:metrics}. A heatmap in Figure~\ref{fig:pca} visualizes these loadings, aiding in identifying key features and optimizing model performance and interoperability.

\subsection{ML Model Development}
\vspace{-0.5cm} 

\begin{table}[]
\centering
\caption{Comparison of Different ML Algorithms.}
\label{tab:pipeline_comparison}
\resizebox{\columnwidth}{!}{%
\begin{tabular}{|l|l|l|l|l|l|l|}
\hline
\textbf{Classifier Name} & \textbf{XGBoost} & \textbf{LightGBM} & \textbf{Random Forest} & \textbf{Extra Trees} & \textbf{Elastic Net} & \textbf{Logistic Regression} \\ \hline
\textbf{Mean CV score} & 0.931 & 0.921 & 0.850 & 0.783 & 0.715 & 0.706 \\ \hline
\end{tabular}%
}
\vspace{-1.5cm}
\end{table}
\begin{table}[h]
    \centering
    \caption{Classification Report of XGBoost model.}
    \begin{tabular}{|l|>{\raggedright\arraybackslash}p{1.5cm}|>{\raggedright\arraybackslash}p{1.6cm}|>{\raggedright\arraybackslash}p{2cm}|>{\raggedright\arraybackslash}p{1.5cm}|}
        \hline
        \multirow{2}{4em}{Class} & \multicolumn{3}{c|}{Metrics} & \multirow{2}{4em}{Support} \\
        \cline{2-4}
         & Precision & Recall & F1-Score & \\
        \hline
        Low & 0.99 & 1.00 & 0.99 & 871 \\
        High & 0.97 & 0.90 & 0.93 & 109 \\
        \hline
        \multicolumn{5}{|c|}{Accuracy: 0.99} \\
        \hline
    \end{tabular}
    
    \label{tab:classification_report}
    \vspace{-0.5cm} 
\end{table}

An ML-based classification model was developed to predict the probability of performance degradation due to interference. The model was trained on an extensive dataset containing various job combinations and their corresponding HC data. Several algorithms were evaluated as shown in Table~\ref{tab:pipeline_comparison}. The XGBoost model was ultimately selected for its superior accuracy and interpretability, achieving a mean cross-validation (CV) score of 0.931 as shown in Table~\ref{tab:classification_report}. The model classifies workloads into two categories: high interference and low interference. High interference refers to workloads experiencing performance degradation of 100\% or more, indicating that the workload performance has fully degraded compared to the non-interference state. Low interference workloads experience degradation below this threshold. The boundary of 100\% degradation was chosen because it represents a critical point where performance has been halved, significantly impacting system efficiency. This threshold distinguishes between workloads that suffer substantial slowdowns due to interference and those that are minimally affected. The high CV score indicates the robustness and effectiveness of the model in predicting performance interference in the SX-AT system.

\subsection{Comparison Between Proposed Approach and Baseline Approach}
\vspace{-0.8cm}

\begin{figure}[ht]
    \centering
    \begin{subfigure}[t]{0.4\textwidth}
        \centering
        \includegraphics[width=\textwidth]{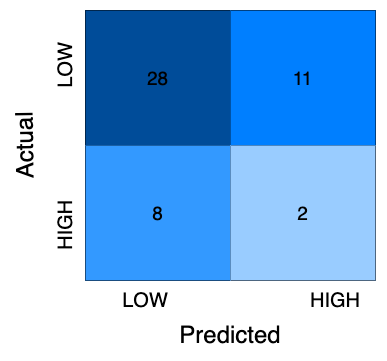} 
        \caption[Baseline Approach]{Baseline Approach.}
        \label{fig:sub1}
    \end{subfigure}
    \hspace{0.05\textwidth} 
    \begin{subfigure}[t]{0.4\textwidth}
        \centering
        \includegraphics[width=\textwidth]{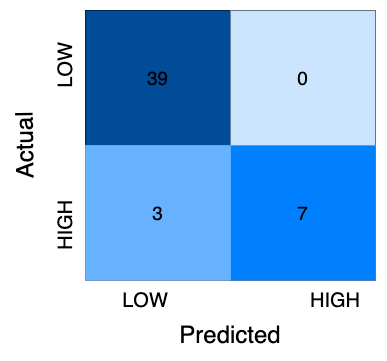} 
        \caption[Proposed Approach]{Proposed Approach.}
        \label{fig:sub2}
    \end{subfigure}
    \caption{Confusion Matrix of Baseline and Proposed Approach.}
    \label{fig:proposed}
    \vspace{-0.5cm}
\end{figure}

\begin{table}[ht]
\centering
\caption{Performance Comparison of Baseline and Proposed Approaches for Unseen Workloads.}
\label{tab:cp}
\vspace{-1.4em}
\begin{tabular}{cc} 
    \begin{subtable}[t]{0.48\linewidth}
        \centering
        \caption{Baseline Approach}
        \label{tab:con_baseline}
        \resizebox{\columnwidth}{!}{%
        \begin{tabular}{|l|l|l|l|l|}
        \hline
        \multirow{2}{*}{Class} & \multicolumn{3}{c|}{Metrics} & \multirow{2}{*}{Support} \\ \cline{2-4}
         & Precision & Recall & F1-Score &  \\ \hline
        Low  & 0.78 & 0.72 & 0.75 & 39 \\ \hline
        High & 0.15 & 0.20 & 0.17 & 10 \\ \hline
        \multicolumn{4}{|c|}{Accuracy: 0.61} & \\ \hline
        \end{tabular}
        }
    \end{subtable}%
    \hspace{0.5cm} 
    \begin{subtable}[t]{0.48\linewidth}
        \centering
        \caption{Proposed Approach}
        \label{tab:con_proposed}
        \resizebox{\columnwidth}{!}{%
        \begin{tabular}{|l|l|l|l|l|}
        \hline
        \multirow{2}{*}{Class} & \multicolumn{3}{c|}{Metrics} & \multirow{2}{*}{Support} \\ \cline{2-4}
         & Precision & Recall & F1-Score &  \\ \hline
        Low  & 0.93 & 1.00 & 0.96 & 39 \\ \hline
        High & 1.00 & 0.7  & 0.82 & 10 \\ \hline
        \multicolumn{4}{|c|}{Accuracy: 0.93} & \\ \hline
        \end{tabular}
        }
    \end{subtable}
\end{tabular}
\vspace{-2em}
\end{table}
This section evaluates the performance of our proposed ML-based interference prediction model against a baseline approach~\cite{nuno}. The baseline method, which relies solely on CPU utilization metrics, achieved moderate accuracy in predicting workload interference, as shown in Table~\ref{tab:con_baseline} and Figure~\ref{fig:sub1}. However, our proposed approach, leveraging a comprehensive set of HC metrics and advanced feature engineering, demonstrated superior performance.

The confusion matrix for our model in Figure~\ref{fig:sub2} reveals significantly improved classification accuracy across all categories of workload interference. Table~\ref{tab:con_proposed} further illustrates the enhanced predictive capabilities of our approach, particularly in identifying high-interference scenarios.
To rigorously evaluate both approaches, we generated a distinct set of unseen workloads, separate from the training and test datasets, to provide a more accurate measure of generalization performance. Our comparative analysis underscores the limitations of CPU-centric metrics in capturing the intricate interactions within the SX-AT system. By integrating diverse performance indicators, our model offers a more comprehensive and nuanced understanding of resource contention and its impact on overall system performance.
These results underscore the potential of our proposed approach in optimizing resource allocation and job scheduling in heterogeneous supercomputing environments, offering a substantial improvement over traditional methods.

\section{Conclusions and Future Work}
This paper presented an ML-based approach to predict performance degradation due to workload interference in the NEC SX-AT vector supercomputer. By leveraging HCs and employing an XGBoost model, the study demonstrated a significant improvement in predicting interference over traditional methods that rely on single metrics such as CPU load. The proposed model achieved a high CV score of 0.931, effectively identifying key features that contribute to performance degradation and enabling more efficient resource allocation. The findings highlight the potential for using ML techniques to dynamically manage system resources, thereby enhancing overall system performance and utilization.

Future research will focus on integrating the predictive model into dynamic job scheduling algorithms to further optimize resource allocation in SX-AT and heterogeneous HPC environments. Additionally, we plan to explore advanced feature selection techniques and deep learning models to capture more complex patterns of interference. Expanding the dataset with more diverse workloads and system configurations will also be considered to improve model generalization. Furthermore, the development of a feedback mechanism that dynamically adjusts model parameters based on real-time performance data will be investigated to enhance the adaptability and effectiveness of the model in evolving HPC settings.
%
%
\vspace{-0.3cm}
\bibliographystyle{splncs04}
\bibliography{ref.bib}

\begin{thebibliography}{10}
\providecommand{\url}[1]{\texttt{#1}}
\providecommand{\urlprefix}{URL }
\providecommand{\doi}[1]{https://doi.org/#1}

\bibitem{ior}
{IOR Benchmark Repository}. \url{https://github.com/hpc/ior}, accessed: 07/29/2024

\bibitem{himeno}
Himeno, R.: {Himeno benchmark }. \url{https://i.riken.jp/en/supercom/documents/himenobmt/}, accessed: 07/29/2024

\bibitem{pcm}
{Intel Corporation}: Intel performance counter monitor. \url{https://github.com/intel/pcm}, accessed: 07/29/2024

\bibitem{intel}
{Intel Corporation}: {Intel MPI Benchmarks}. \url{https://www.intel.com/content/www/us/en/docs/mpi-library/user-guide-benchmarks/2021-2/overview.html} (2021), accessed: 07/29/2024

\bibitem{ke}
Ke, Y., Agung, M., Takizawa, H.: {neoSYCL: a SYCL implementation for SX-Aurora TSUBASA}. In: The International Conference on High Performance Computing in Asia-Pacific Region. p. 50–57. HPCAsia '21, Association for Computing Machinery, New York, NY, USA (2021)

\bibitem{Komatsu}
Komatsu, K., Momose, S., Isobe, Y., Watanabe, O., Musa, A., Yokokawa, M., Aoyama, T., Sato, M., Kobayashi, H.: {Performance Evaluation of a Vector Supercomputer SX-Aurora TSUBASA}. In: SC18: International Conference for High-Performance Computing, Networking, Storage and Analysis. pp. 685--696 (2018)

\bibitem{stream}
McCalpin, J.D.: {Memory Bandwidth and Machine Balance in Current High Performance Computers}. IEEE Computer Society Technical Committee on Computer Architecture (TCCA) Newsletter pp. 19--25 (Dec 1995)

\bibitem{sx}
{NEC Corporation}: {SX-Aurora~TSUBASA Architecture Guide}. \url{https://sxauroratsubasa.sakura.ne.jp/documents/guide/pdfs/Aurora_ISA_guide.pdf} (2024), accessed: 2024-07-29

\bibitem{nuno}
Nunokawa, R., Shimomura, Y., Agung, M., Egawa, R., Takizawa, H.: {Conflict-aware workload co-execution on SX-Aurora TSUBASA}. CCF Transactions on High Performance Computing pp. 1--14 (2023)

\bibitem{hpl}
Petitet, A., Whaley, R.C., Dongarra, J., Cleary, A.: {HPL} - {A} {Portable} {Implementation} of the {High}-{Performance} {Linpack} {Benchmark} for {Distributed}-{Memory} {Computers}. \url{https://www.netlib.org/benchmark/hpl/} (2021), version 2.3

\bibitem{b_eff}
Rabenseifner, R., Hlrs, Koniges, A.E.: {The Parallel Communication and I/O Bandwidth Benchmarks: b\_eff and b\_eff io}. \url{https://api.semanticscholar.org/CorpusID:16156858} (2001), accessed: 07/29/2024

\bibitem{yuta}
Sasaki, Y., Ishizuka, A., Agung, M., Takizawa, H.: {Evaluating I/O Acceleration Mechanisms of SX-Aurora TSUBASA}. In: 2021 IEEE International Parallel and Distributed Processing Symposium Workshops (IPDPSW). pp. 752--759 (2021)

\bibitem{mini}
Sasidharan, A., Snir, M.: {MiniAMR - A miniapp for Adaptive Mesh Refinement}. Tech. rep., University of Illinois (2016)

\bibitem{takahashi}
Takahashi, K., Fujimoto, S., Nagase, S., Isobe, Y., Shimomura, Y., Egawa, R., Takizawa, H.: {Performance Evaluation of a Next-Generation SX-Aurora TSUBASA Vector Supercomputer}. In: High Performance Computing. Springer Nature Switzerland (2023)

\bibitem{takizawa}
Takizawa, H., Shiotsuki, S., Ebata, N., Egawa, R.: {OpenCL-like offloading with metaprogramming for SX-Aurora TSUBASA}. Parallel Computing  \textbf{102},  102754 (2021)

\bibitem{yamada}
Yamada, Y., Momose, S.: {Vector engine processor of NEC’s brand-new supercomputer SX-Aurora TSUBASA}. In: International Symposium on High-Performance Chips (Hot Chips2018) (2018)

\end{thebibliography}
%
\end{document}